# The Theory of Matter in Indian Physics

Roopa Hulikal Narayan

**Abstract**

This paper is the second in series of the Indian physics of the Nyaya-Vaisheshika school. It may be read in conjunction with the first paper [14], where its concept of matter as vibratory atoms in combination was introduced; this concept is discussed in greater detail in this paper. It is significant that the school defines matter not in terms of something gross that is anchored to the commonsensical notion of an object, but rather in terms of something that has attributes associated with it. Matter, or *padartha*, is whatever is knowable within the overarching complex of space and time, each of which is taken to be continuous and infinite. The significant concepts discussed here include how to localize an object, and that of nothingness, that is, vacuum.

**1 Introduction**

This paper presents the physics related to matter in the Indian Nyaya-Vaisheshika school, which will henceforth be called "Indian physics". Kanada, the originator of Vaisheshika, begins by claiming that "classification of things" is the primary task in his system. The purpose of this classification is to define material things. The understanding is to be arrived at using '*tattva-jnana*' [15], or ascertainment of attributes of reality by categorizing everything in to one or the other among six predicable *padartha*, logical categories [3].

Indian physics considers both the objective universe, which is taken to be atomic, and the subjective universe of the experimenter or the observer, which is taken to be non-atomic [7], [[11], and [12]. In other words, it presents a dualistic view where that the observed matter is atomic whereas the observing mind and time and space in which the universe exists is continuous.

**2. Categories and localization**

We begin by noting that the concept of '*tattva-jnana*' or 'ascertainment of reality' is the principle that there is no thought (or conceptual framework) that cannot be expressed [15]. Everything that can be stated belongs to the highest class of knowledge called *padartha*, or "predicable", which is describable and hence nameable. Such nameables, classified to six categories, form the basis of physics.

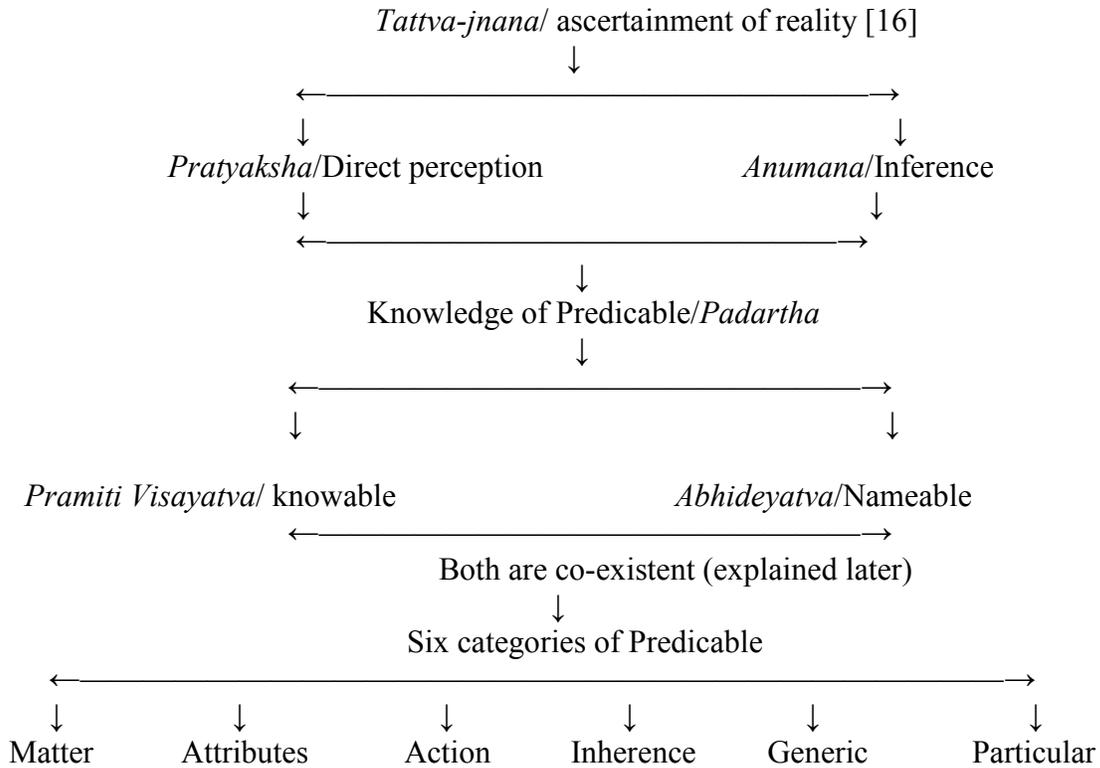

Diagram: 1

Matter is something in which attributes inhere that is associated with action which in its widest class is generic and in the smallest is particular [15]. This broad definition seems to correspond with the commonsensical idea of matter, where it is something situated in space and time that has certain attributes associated with it. The difference is that in addition to whatever attributes one may associate with the object and its action, there are further categories associated with the relationships that the object can enter with other objects.

The purpose of this paper and those to follow is to look carefully at these categories and examine their implications. Before we proceed, we wish to note that in the Indian tradition the cosmos was infinite with a postulated connection between the inner and the outer while in ancient Greek thought that is the foundation of the current western science the universe was a finite system [4], [5]. These differences in the underlying cosmology find expression in the way the two cultures dealt with scientific problems. It may also be noted that the Indian tradition used careful definition of linguistic terms, that were subject to philosophical analysis and, therefore, the tradition was somewhat in the manner of physics as "natural philosophy".

**2.1 Localizing the object**

Vatsyayana, a later Indian philosopher, defines the conditions under which the object to be studied can be differentiated from the rest [2]. He defined the conditions under which it is validly localized using the Sanskrit term लक्षण (*lakshana*) – the distinguishing features to define an entity [6] – that is essential for its proper definitis.

Although this might appear a linguistic question associated with '*vishesha*" (particularity), this issue is of importance in the formulation of the conceptual framework underlying the physics. Rather than see localization in terms of a mental map related to an object having been physically isolated from others, the school uses three categories to see if such localization is adequate:

1:अव्याप्ति (*Avyapti*) under-definition
This definition is too narrow; it excludes a similar object in other circumstances.

2:अतिव्याप्ति (*Athivyapti)* – over-inclusion
This definition is too wide; it includes other objects that are, in reality, different.

Step 3:असंभव –*Asambhava*- impossible
This deals with a definition that does not correspond to any known physical system [6].

These definitions leave open the possibilities that the particularity may have mutually exclusive attributes at the same time as long as they correspond with observation. What is impossible is not defined in an *a priori* manner but on the basis of properties that have been observed. In other words, it gives priority not to any logical description of the system but rather to observations associated with it.

It is quite clear that this approach does not correspond to the realist position of physics; in contemporary terminology this viewpoint seems to be somewhat like the positivist viewpoint [17].

**3 The six categories**

The predicable with the six divisions is based on logical analysis, and since it does not actually specify the particularity, it has the capacity to include new attributes. The term '*padartha*' for predicable means that which can be named or described through a word, or something that can be comprehended by senses. *Padartha* or predicable is interchangeable with *knowable* – that which can be known [2].

It is essential to note the possibility of faulty perception. An object of observation by its common property may seem similar to another existing object as a result of faulty perception or inference. This may be avoided by individuating each *padartha* with *vishesha* (particularity). At the same time, objects of the same genus are classified under the same '*samanya*' (generic).

Kanada declares *samanya* (generic) and *vishesha* (particularity) as names that arise relative to observer. These are, therefore, observer-dependent notions that add to the definition of matter, in terms of attribute and action alone. The last *padartha*, inherence (*samavaya)* is relevant in the context of cause and effect [8].

The emergence of the six-fold division of *padarthas* may be diagrammed as follows:

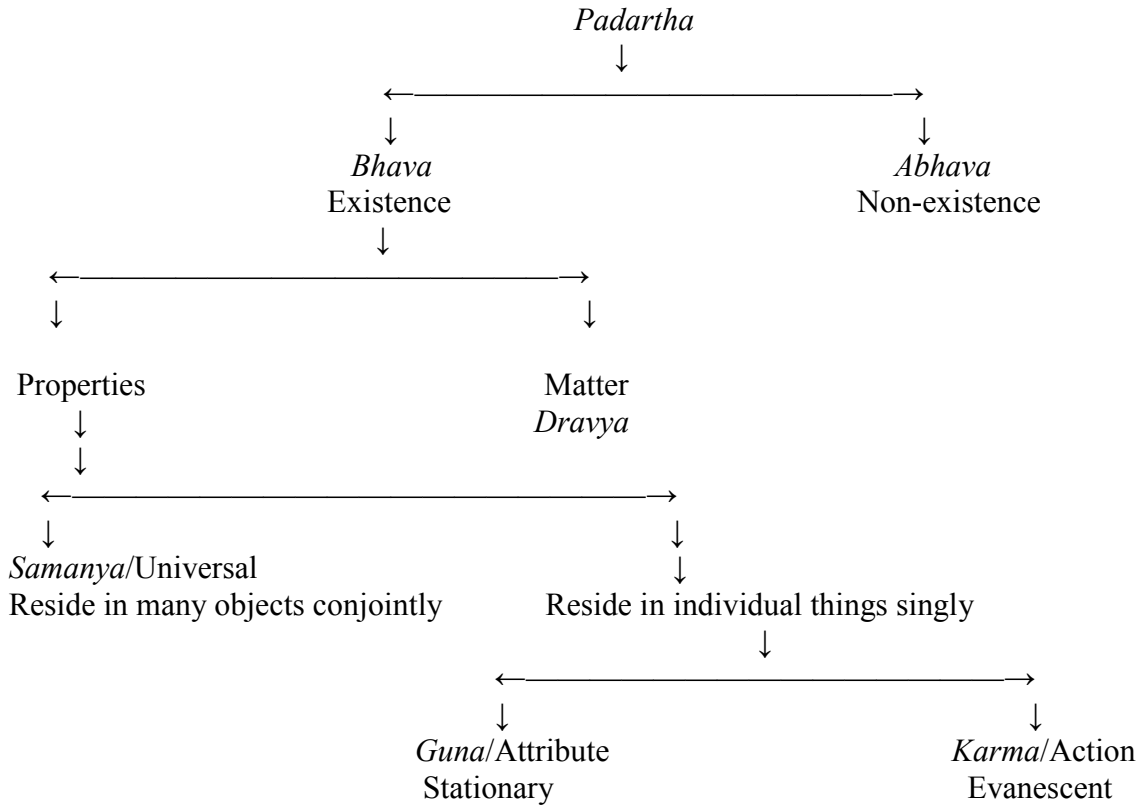

Diagram: 2

*Samavaya* (inherence) and *vishesha* (particularity) are special categories to explain the world.

**3.1 Vacuum -- the seventh predicable**

Although a*bhava* (non- existence or nothingness) was not included by Kanada as a category, later thinkers argued the need to recognize 'denial of a known' as a category by itself [2]. This was a category like that of vacuum in medieval physics, and we will, now, use this term for it rather than the cognitively jarring "non-existence" that has generally been used to translate it.

An example to understand the perspective of vacuum as a category is as follows:

Statement: 'Vacuum' is a form of matter.

*Argument from direct consideration***:**

1. Vacuum must be classified as matter because it possesses attributes (*guna*) that no other known matter/*dravya* possesses.
2. Color excludes vacuum from belonging to the category of ether, space, time, and so on, all of which are colorless.
3. It cannot be air as it lacks 'touch' and 'constant motion' – the attributes of air.
4. It is not light as it lacks brightness and hot touch.
5. It is not water for it lacks cool touch or water color.
6. It is not earth as it lacks scent or touch.
7. Therefore it is a new kind of matter not belonging to any of the existing categories.

*Argument related to absence:*

1. Matter is perceived when light reflects off it.
2. Therefore vacuum is perceived only in the absence of all kinds of light.

Vacuum is consequently defined as प्रौढप्रकाशकतेजःसामान्याभावः, i.e. the non-existence of light particles at the smallest particle level, not referring to the absence of sources of light like sun, moon or a lamp. This is to emphasize that vacuum is not absence of source of light, instead the very absence of light at the minutest level perceived by mind.

Thus vacuum is an entity. Absence of an entity is cognized only when its presence is recognizable which implies recognition of absence is not possible independently. But opinion was not uniform on this matter in all schools. In the Mimamsa School, vacuum was taken as an independent entity.

**4 Action**

Action or motion (*kriya*) associated with matter is thus defined:

उत्क्षेपणमवक्षेपणमाकुञ्चनं प्रसारणं गमनमिति कर्म्माणि ॥१ । १ । ७ ॥

Action is throwing upwards, downwards, contraction, expansion and general motion. 1.1.7 [1]

The general motion implies rotation as well. Here action (*karma*) refers to the state of 'an action taking place' as understood by the observer. The five independent kinds of motion or states of action by inter-combining produce several other kinds of motions. But all motions can be resolved in to these five basic types of motion.

**4 Attribute (*Guna*)**

द्रव्याश्रय्यगुणवान् संयोगविभागेष्वकारणमनपेक्ष इति गुणलक्षणम् ॥१।१।१६॥

An attribute (*guna*) inheres in matter, it being attribute-less. But it is never an independent cause of conjunction or disjunction where conjunction and disjunction are two kinds of motion as defined before.1.1.6 [1]

The karma/action differs in the fact that it is an independent cause of conjunctions and disjunctions. Action can reside only in one given matter.

**5 Matter differentiated from its properties**

A need to distinguish matter from its attributes or properties in general leads to the theoretical assumption that at the immediate moment of creation matter is attribute-less but possesses 'matter-ness' which is the cause of its matter form. The paradox here is that if all the definable properties of matter are intimately connected with it so as to exist in it, then how is matter different from its properties? On the other hand, if the properties are not those which happen to be part of matter, how can they be used to define or understand it?

**5.1 Explanation:**

क्रियागुणवत् समवायिकारणमिति द्रव्यलक्षणम् ॥१।१।१५॥6

Matter is the combinative cause of *kriya*/action and *guna*/attribute. 1.1.5 [1]

Matter is the inherent cause of action/*kriya* and *guna*/attribute as stated by the above rule. It is expressed and understood as matter through its properties. But, matter comes in to existence first as the antecedent cause, followed by properties such as *kriya*/motion and *guna*/attribute. Therefore there essentially is a moment when matter is attribute-less.

*Coexistence of qualities*
In the absence of a materialistic substratum, different properties will have to be attributed to the same coordinates of space. But these qualities can be perceived only one after another by the very nature of mind, which can deal with information only in a linear sense [17], which means they will be assigned different temporal coordinates in our representation. Therefore the mind perceiving the two qualities together can only be as a result of a common substratum which in reality holds them together [6].

**5.2 Matter and self-subsistence/ *Dravyatvajathi***

In response to arguments against the separation of matter from its properties, Sridhara came up with an interesting theory. Realizing that there is no self-evident commonality among defined substances, he came up with the concept of *Dravyatvajati* (matter-hood).

He describes matter as:

1. It is self-subsistent with a synthetic concept of matter-hood to which all specific matters can be affiliated on account of their possession of some fundamental common character.
2. Matter-hood exists independently unlike *guna*/attribute or *karma*/action which makes sense only in relation to a matter.
3. Matter-hood cannot be independent of the matter's existence since in material-bodies the whole and parts are mutually dependent to exist as a composite whole matter.
4. Matter-hood cannot be perceived independently of matter because for e.g. 'touch of air' is perceived even in the absence of its substrata air not being perceived.

Further issues:

5. Two types of matters exist: (i) eternal and (ii) non-eternal.

6. The atoms, time, space, etc which are eternal are self-subsistent by nature because their eternality translates in to an existence by their own right.

7. Material-bodies which are non-eternal possess no such absolute self-subsistence because they constitute of parts in which they inhere. They do possess a relative self-subsistence with respect to the attributes and action that inhere in them [6].

**6 Matter, motion (*kriya*) and attribute (*guna*)**

By the sutra 1.1.5 both action and attributes inhere in matter. Does this imply that action and attribute are the necessary or sufficient conditions of matter being matter? This is considered as follows.

**6.1 *Kriya* (Motion)**

*Kriya* (motion) is by its very nature restricted to the eternal entities which are incapable of any physical or material motion. Therefore, *kriya* (action) is a narrow definition of matter, i.e. it corresponds to the case of *avyapti* (definition is too narrow).

Sankara Misra defends by stating that *kriya* (action) necessarily implies existence of matter but not the contrary.

Jayanarayana has two interpretations:

1. *Kriya*/action inheres in all that is capable of motion which is matter alone. The eternal entities are not excluded since they share a common generic character of *dravyatvajati* (matter-hood) with the non-eternal entity.

2. *Kriya*/action inheres in the eternal matter indirectly as in *akasha* being a substratum of conjunction and disjunction wherein the materialistic motion is performed by the actual atoms which conjunct or disunite [6].

## 6.2 *Kriya*/motion for non-material entities

The idea of motion is generally understood as the physical motion of a material body. This is a very limited definition when non-material kinds of matter are considered as existing and affecting the world. The motion of non-material entities can't be measured by the material body units. This is the equivalent of measuring spherical space with two-dimensional co-ordinates. Further, note that mind travels through thought or the very thought is mind which can travel. Therefore by expanding the definition of *kriya* mind too is a substratum for it.

## 6.3 *Akasha*/Ether

The difference between *akasha* and space is the attribute/*guna* of sound inhering in *akasha* and not in space. The two are same for all other considerations. Sound was understood as propagating in wave form similar to water ripples through a propagating medium [Page 16, 5] but light was treated as particles [13, Page 105]. All the same if sound requires a medium of propagation and therefore *akasha* is that which contains a medium, such a medium has to be necessarily capable of *kriya*/motion by the very reason of sound propagation.

## 6.4 Time

Time, having been considered infinite like space, cannot be defined in terms of events because that would be subjective. But what is the relevance of time if there is nothing that needs to be measured? Kanada's sutra about time states

द्रव्यत्वनित्यत्वे वायुना व्याख्याते ॥२।२।७॥
The matter-ness and eternality (of time) are explained by (the explanation of the matter-ness and eternality of) air.  2.2.7 [1]

Time is compared to the atom of air and is treated as a matter in the same spirit. In the next sutra Kanada defines time as the cause of everything produced or created. Matter and other things by themselves could not be the cause of these notions of entities as these are wholly different from the notions of matter &c; nor could any effect be produced without an adequate cause; hence we conclude that that which is the cause of these is time.

Argument:
1. If time is understood as past, present and future it is merely circumstantial. How can it be causal?

Reply by Vaisheshikas:

1. 'Producing 'is only for a previously non-existing object. In the absence of time no such qualifier for before and after being available, an object produced will not be any different from an *akasha* – an absolute existence or a human with horns- an absolute non-existence.
2. Time is causal for a qualified cognition in the sense of a sense of organ which helps identification [9].

In such a definition of time, what is meant by 'causal' for a sense organ needs to be understood.

Time is absolute and should have an existence independent of the observer by Vaisheshika's definition of time. Kanada describes time as *asamavayi karana* i.e. non-inherent cause but Prashastapada describes it as *nimmitta karana*-efficient cause. Time is described as recognized by its effects and so also space. In modern sense the two are treated as one in reality [10, Page 136, 137].

In a later sutra about time, Kanada states

नित्येष्वभावादनित्येषु भावात् कारणे कालाख्येति॥२।२।६॥

The name time is applicable to a cause in as much as it does not exist in eternal matters and exists in non-eternal matters. 2.2.6

This is explained as in eternal matter which is neither created nor destroyed by its very nature, the concept of 'produced simultaneously', 'produced quickly or early', 'produced in day or night', etc is not applicable. In non-eternal matter which has a beginning and end such concepts are applicable as explained above.

**6.6 *Guna* (Attribute)**

Is attribute a necessary condition for matter? If so how does matter continue to remain so at the moment of its formation when it is attribute-less? This is explained by Vallabha, another philosopher of Vaisheshika School, considering matter as a potential substratum of attribute even at the time it is attribute-less. Therefore *Gunavat* (inherence) of attribute (sutra 1.1.5) is interpreted to mean even potential existence of attribute in it as in the attribute-less state [2]. A similar problem does not arise with *kriya*/action because all material bodies have vibratory motion (*parispanda)* of atoms in all states in Vaisheshika.

An attribute need never be confused for matter because it is not attributable or cannot be the substratum for another attribute and so can never be a matter.

**7 Matter and Properties**

The following table compares the only nine different kinds of matter of Vaisheshika in terms of their properties.

Table 1 [10]

| Matter (with/which is) | Earth | Water | Fire | Wind | Mind | Ether | Space | Time | Self |
|---|---|---|---|---|---|---|---|---|---|
| 1. Active | * | * | * | * | * | – | – | – | – |
| 2. Attributes | * | * | * | * | * | * | * | * | * |
| 3. Touch | * | * | * | * | – | – | – | – | – |
| 4. Color | * | * | * | – | – | – | – | – | – |
| 5. Eternality | – | – | – | – | * | * | * | * | * |

Table: 1

## 8 *Abhava*

We now consider a*bhava,* normally viewed as nothingness or non-existence, which is one of the most remarkable and surprising contributions of Indian physical thought.

8.1 *Abhava* as negation, a logical category

*Abhava* as negation is a means of inference in four ways [2] diagrammed below:

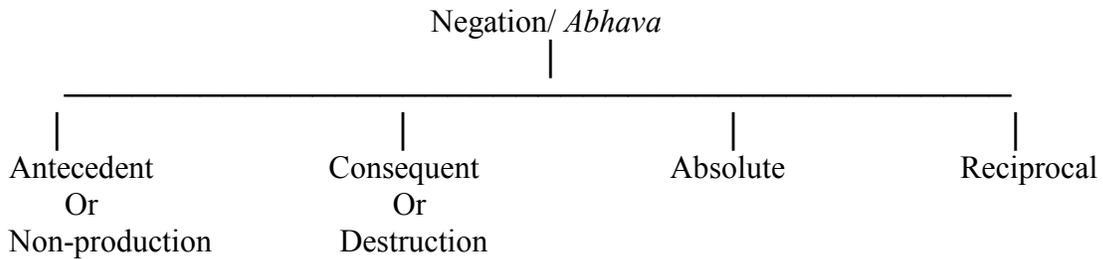

Diagram: 3

Antecedent Negation is that which exists before a thing is produced and Consequent is what follows after the destruction of a thing.

Reciprocal Negation is the denial of one thing being another which is helpful in defining a thing as e.g. a computer cannot be a clock because it is more complex, etc.

Absolute Negation is what interests us. It is defined as that which exists always and in all places except where the thing itself is.

Visvanatha in his Bhasa-Pariccheda [2] classifies absolute *abhava* under a sub-classification of संसर्गाभाव which literally means 'negation by contact' where the contact is between the thing negated and the thing on which the negation is affirmed.

This above definition when applied to absolute *abhava* can be understood as mutual negation between two things as follows:

There is no matter nature in absolute *abhava*
There is no absolute *abhava* in matter.

Further, absolute *abhava* is defined as that which has no volition, impulse or impact, gravity or fluidity, and no resultant energy [1, Page 186]. This definition of absolute *abhava* is nothing but the definition of vacuum.

**8.2 *Abhava* as Vacuum**

The term *astitva* (being-ness) is a positive cognition which is the capability of an object being perceived independently of counter-entity [9, Page 37] where as 'being a predicable' is the capability of being expressed in words. In Kiranavali a later work, it is admitted that both *astitva* and being predicable belong to negation as well though not included in the six defined categories of Kanada which does not restrict their presence elsewhere [9].

The existence of absolute *abhava* is not denied though it is denied as a category of matter by Kanada owing to its dependency on the absence of light where the dependency weighs heavy for this specific kind of positive cognition based enumeration of predicable. "Predicable" the translation of '*Padartha*' as used by Kanada is accurate in that the definition of the term predicable in logic in English is 'affirmation or denial about something'.

Dependency is defined as a character of all non-eternal things which excludes time, space, etc. Also the character of being an effect and that of being non-eternal belong only to those that have causes which are matter, action and attribute [9].

The entire universe is non-eternal in the sense that it is created and will be destroyed which means it is an effect. Can this mean that the other universe and the universe which we perceive are mutual causes of each other and by the very nature of the instrument of perception which is the observer in one universe at a time, each universe is the negation of the other or the *abhava*/vacuum for the other?

**8.2 *Samavaya* as related to Vacuum**

The above stated point becomes important when further the idea of *samavaya* (inherence) is examined. This is a key concept of Kanada which etymologically means an 'intimate union' between two things which are rendered inseparable so that they cannot be separated without themselves being destroyed [2].

But time is causal as explained in a previous section on time and *samavaya* is defined as a relationship between the cause and its effects whereby the one is cognized as residing in the other [9]. In an experiment where observations appear to reduce the physical state [17] time is inhered in the observer even though Kanada defines time as *asamvayi karana* – or the non-inhering cause. Kanada who was material – matter oriented probably was focusing only on inherence in atomic matter rather than an observer who is mind + consciousness + intelligence, etc. This probably means that time is inhered in the observer in an inherent manner which means it is not separable from the observer nor is it something that is learnt but just a part of her.

**9 Conclusions**

The basic Vaisheshika concepts of matter, action, attribute, time and space are described. Matter, consisting of vibratory atoms, can be localized by its attributes that need not correspond to any straight-forward physical map. This view is somewhat like that of the positivists.

It is significant that the school defines matter not in terms of something gross that is anchored to the commonsensical notion of an object, but rather in terms of something that has attributes associated with it. Matter or *padartha*, is whatever is knowable within the framework of the overarching spacetime, which is taken to be continuous and infinite.

The most surprising aspect of this conception is the idea of "nothingness" or "vacuum". It is defined as a unique category in itself, subject to analysis by determining its relationship to the observer.